\documentclass[sigchi]{acmart}

\usepackage{balance}
\usepackage{booktabs} 
\usepackage{multirow}
\usepackage{color}
\usepackage[colorinlistoftodos]{todonotes}
\usepackage{enumitem}
\usepackage{boldline}
\usepackage{ctable}

\copyrightyear{2019} 
\acmYear{2019} 
\setcopyright{acmlicensed}
\acmConference[CHI 2019]{CHI Conference on Human Factors in Computing Systems Proceedings}{May 4--9, 2019}{Glasgow, Scotland Uk}
\acmBooktitle{CHI Conference on Human Factors in Computing Systems Proceedings (CHI 2019), May 4--9, 2019, Glasgow, Scotland Uk}
\acmPrice{15.00}
\acmDOI{10.1145/3290605.3300305}
\acmISBN{978-1-4503-5970-2/19/05}

\begin{document}
\title[Mobile Interface Tappability Modeling]{Modeling Mobile Interface Tappability Using Crowdsourcing and Deep Learning}


\author{Amanda Swearngin}
\authornote{This work was completed while the author was an intern at Google.}
\affiliation{
  \institution{University of Washington}
  \city{Seattle}
  \state{WA}
}

\email{amaswea@cs.washington.edu}
\author{Yang Li}
\affiliation{%
  \institution{Google Research}
  \city{Mountain View}
  \state{CA}
}
\email{yangli@acm.org}

\begin{abstract}
Tapping is an immensely important gesture in mobile touchscreen interfaces, yet people still frequently are required to learn which elements are tappable through trial and error. Predicting human behavior for this everyday gesture can help mobile app designers understand an important aspect of the usability of their apps without having to run a user study. In this paper, we present an approach for modeling tappability of mobile interfaces at scale. We conducted large-scale data  collection of interface tappability over a rich set of mobile apps using crowdsourcing and computationally investigated a variety of signifiers that people use to distinguish tappable versus not-tappable elements. Based on the dataset, we developed and trained a deep neural network that predicts how likely a user will perceive an interface element as tappable versus not tappable. Using the trained tappability model, we developed TapShoe, a tool that automatically diagnoses mismatches between the tappability of each element as perceived by a human user---predicted by our model, and the intended or actual tappable state of the element specified by the developer or designer. Our model achieved reasonable accuracy: mean precision 90.2\% and recall 87.0\%, in matching human perception on identifying tappable UI elements. The tappability model and TapShoe were well received by designers via an informal evaluation with 7 professional interaction designers.
\end{abstract}

%
%

\ccsdesc[500]{Human-centered computing~Interactive systems and tools}
\ccsdesc[500]{Human-centered computing~Systems and tools for interaction design}

\keywords{Mobile interfaces; tappability, crowdsourcing, deep learning;}

\begin{teaserfigure}
  \includegraphics[width=\columnwidth]{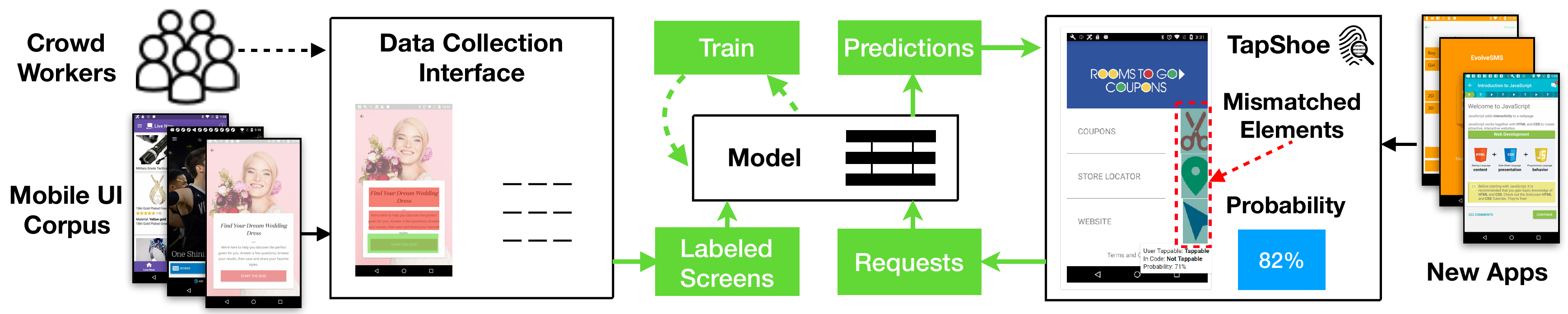}
  \caption{Our deep model learns from a large-scale dataset of mobile tappability collected via crowdsourcing. It predicts tappability of interface elements and identifies mismatches between designer intention and user perception, and is served in the TapShoe tool that can help designers and developers to uncover potential usability issues about their mobile interfaces.}
  \Description{The figure displays a diagram of our approach to modeling tappability and presenting results of the model in the TapShoe interface. The data collection process takes in a large set of mobile UIs, each of which is labeled in a data collection interface by crowdworkers. The interface lets the workers label interface elements as tappable or not tappable. The labeled screens and interface elements are then passed into a deep learning model for training. The model learns from the features of the elements and the labels, and produces predictions. The TapShoe interface is a prototype interface that queries the model for predictions. Designers and developers can upload their apps to the TapShoe interface and see predictions for each element and whether the prediction has a mismatch with the actual tappable state of the element in code. 
  }
  \label{fig:teaser}
\end{teaserfigure}

\maketitle

\section{Introduction}
Tapping is arguably the most important gesture on mobile interfaces. Yet, it is still difficult for people to distinguish tappable and not-tappable elements in a mobile interface. In traditional desktop GUIs, the style of clickable elements (e.g., buttons) are often conventionally defined. However, with the diverse styles of mobile interfaces, tappability has become a crucial usability issue. Poor tappability can lead to a lack of discoverability \cite{norman2008signifiers} and false affordances \cite{gaver1991affordances} that can lead to user frustration, uncertainty, and errors \cite{nngroup2015, nngroup2017}. 

Signifiers can \cite{norman2008signifiers} indicate to a user how to interact with an interface element. Designers can use visual properties (e.g., color or depth) to signify an element's "clickability" \cite{nngroup2015} or "tappability" in mobile interfaces. Perhaps the most ubiquitous signifiers in today's interfaces are the blue color and underline of a link, and the design of a button that both strongly signify to the user that they should be clicked. These common signifiers have been learned over time and are well understood to indicate clickability \cite{norman1999Affordances}. To design for tappability, designers can apply existing design guidelines for clickability \cite{nngroup2015}. These are important and can cover typical cases, however, it is not always clear when to apply them in each specific design setting. Frequently, mobile app developers are not equipped with such knowledge. Despite the existence of simple guidelines, we found a significant amount of tappability misperception in real mobile interfaces, as shown in the dataset we discuss later. 

Additionally, modern platforms for mobile apps frequently introduce new design patterns and interface elements.  Designing these to include appropriate signifiers for tappability is challenging. Additionally, mobile interfaces cannot apply useful clickability cues available in web and desktop interfaces (e.g., hover states). With the flat design trend, traditional signifiers have been altered, which potentially causes uncertainty and mistakes \cite{nngroup2017}. More data may be needed to confirm these results, however, we argue that we need more data and automated methods to fully understand the users' perceptions of tappability as design trends evolve over time. 

One way that interface designers can understand tappability in their interfaces is through conducting a tappability study or a visual affordance test \cite{visualAffordanceTesting}. However, it is time-consuming to conduct such studies. In addition, the findings from these studies are often limited to a specific app or interface design. We aim to understand signifiers at a large scale across a diverse array of interface designs and to diagnose tappability problems in new apps automatically without conducting user studies.

In this work, we present an approach for modeling interface tappability at scale. In addition to acquiring a deeper understanding about tappability, we develop tools that can automatically identify tappability issues in a mobile app interface (see Figure~\ref{fig:teaser}). We trained a deep learning model based on a large dataset of labeled tappability of mobile interfaces collected via crowdsourcing. The dataset includes more than 20,000 examples from more than 3,000 mobile screens.  Our tappability model achieved reasonable accuracy with mean precision 90.2\% and recall 87.0\% on identifying tappable elements as perceived by humans. To showcase a potential use of the model, we build TapShoe, a web interface that diagnoses mismatches between the human perception of the tappability of an interface element and its actual state in the interface code. We conducted informal interviews with 7 professional interface designers who were positive about the TapShoe interface, and could envision intriguing uses of the tappability model in realistic design situations. Our contributions include the following:
\begin{enumerate}[noitemsep,topsep=0pt]
    \item An approach for understanding interface tappability at scale using crowdsourcing and computational signifier analysis, and a set of findings about mobile tappability;
    \item A deep neural network model that learns human perceived tappability of interface elements from a range of interface features, including the spatial, semantic and visual aspects of an interface element and its screen, and an in-depth analysis about the model behavior;
    \item An interactive system that uses the model to examine a mobile interface by automatically scanning the tappability of each element on the interface, and identifies mismatches with their intended tappable behavior. 
\end{enumerate}

\section{Related Work}
The concepts of signifiers and affordances are integral to our work. Our aim is to capture them in a structured way to construct a predictive model and to understand their use in a large set of real mobile interfaces. Affordances were originally described by \cite{gibson1978ecological} as the actionable properties between the world and actor (i.e., person). Don Norman \cite{norman1999Affordances, norman2013design} popularized the idea of affordances of everyday objects, such as a door, and later introduced the concept of a "signifier" as it relates to user interfaces \cite{norman1999Affordances}. Gaver \cite{gaver1991affordances} described the use of graphical techniques to aid human perception (e.g., shadows or rounded corners), and showed how designers can use signifiers to convey an interface element's perceived affordances. These early works form the core of our current understanding of what makes a person know what is interactive. By collecting a large dataset of tappability examples, we hope to aid our understanding of which signifiers are having an impact at scale. 

Since those early works, there have been a few studies about the factors influencing clickability in web interfaces \cite{nngroup2015, nngroup2017}. Usability testing methods have also adopted the idea of visual affordance testing \cite{visualAffordanceTesting} to diagnose clickability issues. However, these studies have been conducted at a small scale and are typically limited to the single app being tested. We are not aware of any large-scale data collection and analysis across app interfaces to enable diagnosis of tappability issues, nor any machine learning approaches that learn from this data to automatically predict the elements that users will perceive as tappable or not tappable.

To identify tappability issues automatically, we needed to collect data on a large scale to allow us to use a machine learning approach for this problem. Recently, data-driven approaches have been used to identify usability issues \cite{deka2017zipt}, and collect mobile app design data at scale \cite{deka2016erica, deka2017rico}. Perhaps most closely related to our work is Zipt \cite{deka2017zipt}, which enables comparative user performance testing at scale. Zipt uses crowd workers to construct user flow visualizations through apps that can help designers visualize the paths users will take through their app for specific tasks. However, with this approach, designers must still manually diagnose the usability issues by examining the visualizations. In this paper, we focus on an important usability issue in mobile interfaces automatically---identifying cases where false affordances or missing signifiers will cause a user to misidentify a tappable or not-tappable interface element.

Similar to Zipt \cite{deka2017zipt}, our work uses crowdsourcing to collect user data to aid the diagnosis of usability issues. We used Amazon's Mechanical Turk that has previously provided a platform for large-scale usability \cite{nebeling2013crowdstudy} and human subjects experiments \cite{komarov2013crowd, kittur2008crowd, schneider2016crowdux}, and in gathering data about the visual design of user interfaces \cite{luther15crowdesign, xu2014voyant, greenberg2015crowd}. Our work goes beyond data collection and analysis by developing machine learning models to automatically examine tappability. 

Deep learning \cite{lecun2015deep} is an effective approach to learn from a large-scale dataset. In our work, we trained a deep feedforward network, which uses convolutional layers for image processing and embedding for categorical data such as words and types, to automatically predict human tappability perception. Recent work has used deep learning approaches to predict human performance on mobile apps for tasks such as grid \cite{pfeuffer2018grid} and menu selection \cite{li2018predicting}. Deep learning models have also been built to identify salient elements in graphic designs and interfaces \cite{bylinskii1027saliency}. However, no work has applied these models to predicting the tappability of interface elements. Deep learning allowed us to leverage a rich set of features involving the semantic, spatial and visual properties of an element without extensive feature engineering.

\section{Understanding Tappability at Scale}
A common type of usability testing is a \textit{tappability study} or a visual affordance test \cite{visualAffordanceTesting}. In these studies, designers have crowd workers or lab participants label interfaces for which elements they think are tappable and not tappable digitally or on paper. Based on this data, designers can construct heatmaps to visualize where users would tap in the app being tested. These studies can help designers discover which elements have missing or false tappability signifiers. However, in general, there is a lack of a dataset and deep understanding about interface tappability across diverse mobile apps. Having such a dataset and knowledge is required for us to create automated techniques to help designers diagnose tappability issues in their interfaces.

\begin{figure*}
\centering
\includegraphics[width=0.6\textwidth]{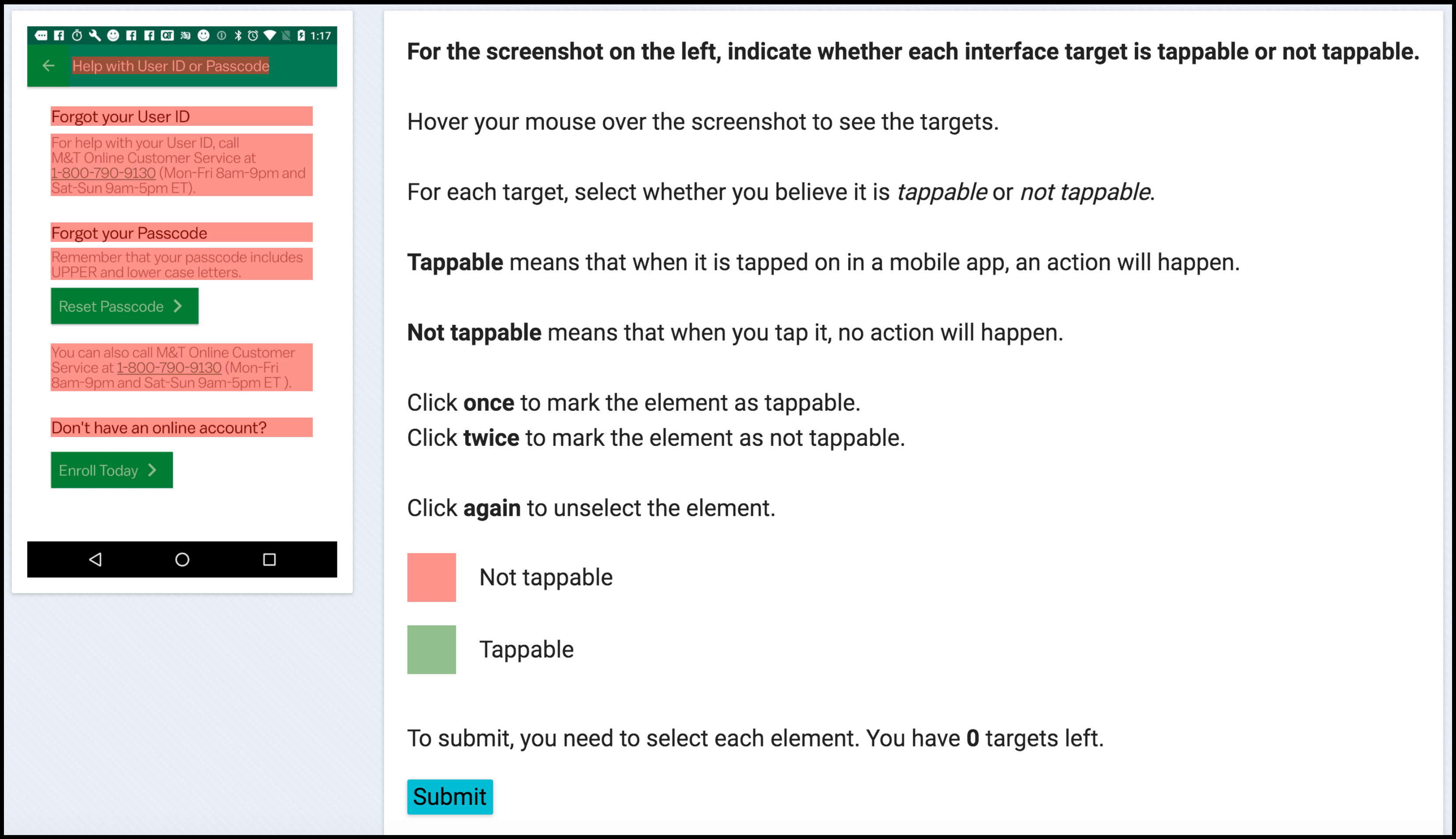}
    \caption{The interface that workers used to label the tappability of UI elements via crowdsourcing. It displays a mobile interface screen with interactive hotspots that can be clicked to label an element as either tappable or not tappable.}
    \Description{The crowdsourcing interface displaying a screenshot of a mobile app on the left hand side, with red and green boxes over various interface elements. The right hand side displays the set of instructions for the task that tell the crowd worker to label each element as tappable or not tappable. The instructions tell the crowd worker to hover their mouse over the screenshot to see grey highlighted hotspots, and click once or twice to mark an element as tappable or not tappable. The interface also has a rubric that displays the colors mapping to tappable or not tappable (i.e., red or green) and a submit button they can click after labeling all highlighted elements.}
    \label{fig:tappability_task}
    \vspace{-1em}
\end{figure*}

\subsection{Crowdsourcing Data Collection}
We designed a crowdsourcing task to simulate a tappability study across a large corpus of Android mobile apps \cite{deka2017rico}, using the interface shown in Figure \ref{fig:tappability_task}. The left side of the interface displayed a mobile app screenshot. The right side of the task interface displayed instructions for the task, and an explanation about what we meant by tappable and not tappable. For tappable elements, it was \textit{"When you tap this in a mobile interface, an action will happen."}, and for not tappable, the explanation was \textit{"When you tap on it, no action will happen."}. 

To collect our tappability dataset, we selected a set of 3,470 unique, randomly chosen screens from the Rico dataset \cite{deka2017rico} and had crowd workers label elements randomly sampled from these screens as either \texttt{tappable} or \texttt{not tappable}. We selected the elements for the workers to label in the following manner. Each UI screen in the Rico dataset has an Android view hierarchy---JSON tree structure of all of the interface elements on the screen, similar to a DOM tree for a web interface. Each element in the hierarchy has a \texttt{clickable} property that marks whether an element will respond to a tapping event. For each screen, we selected up to five unique \texttt{clickable} and non-\texttt{clickable} elements. When selecting \texttt{clickable} elements, starting from a leaf element, we select the top-most \texttt{clickable} element in the hierarchy for labeling. When a clickable element contains a sub-tree of elements, these elements are typically presented as a single interface element to the user, which is more appropriate for the worker to label as a whole. When a clickable container (e.g., ViewGroup) is selected, we do not select any of its child elements thus preventing any duplicate counting or labeling. We did not select elements in the status bar or navigation bar as they are standard across most screens in the dataset.

To perform a labeling task, a crowd worker hovers their mouse over the interface screenshot, and our web interface displays grey hotspots over the interface elements pre-selected based on the above process. Workers click on each hotspot to toggle the label as either tappable or not tappable, which are colored in green and red, respectively. We asked each worker to label around six elements for each screen. Depending on the screen complexity, the amount of elements could vary. We randomized the elements as well as the order to be labeled across each worker.


\begin{table}
\centering
\begin{tabular}{l c c c c}
   & \textbf{Positive Class} & \textbf{\#Elements} & \textbf{Precision} & \textbf{Recall} \\ 
\midrule[\heavyrulewidth]
\parbox[t]{1mm}{\multirow{3}{*}{\rotatebox[origin=c]{90}{\textbf{R1}}}} &  \texttt{clickable=True} & 6,101 &  79.81\% & 89.07\% \\
& \texttt{clickable=False} & 3,631 & 78.56\% & 61.75\%\\
\midrule[\heavyrulewidth]
\parbox[t]{1mm}{\multirow{3}{*}{\rotatebox[origin=c]{90}{\textbf{R2}}}}  & \texttt{clickable=True} & 6,560 & 79.55\%  & 90.02\% \\
& \texttt{clickable=False} & 3,882 & 78.30\% & 60.90\% \\
\midrule[\heavyrulewidth]
\parbox[t]{1mm}{\multirow{3}{*}{\rotatebox[origin=c]{90}{\textbf{All}}}} &
\texttt{clickable=True} & 12,661 & 79.67\% & 89.99\% \\
& \texttt{clickable=False} & 7,513 & 78.43\% & 61.31\% \\
\end{tabular}
\vspace{1em}
\caption{The number of elements labeled by the crowd workers in two rounds, along the precision and recall of human workers in perceiving the actual clickable state of an element as specified in the view hierarchy metadata.}~\label{tab:accuracy_and_labels}
\vspace{-3em}
\end{table}

\subsection{Results}
We collected 20,174 unique interface elements from 3,470 app screens. These elements were labeled by 743 unique workers in two rounds where each round involved different sets of workers (see Table \ref{tab:accuracy_and_labels}). Each worker could complete up to 8 tasks. On average, each worker completed 4.67 tasks. Of these elements, 12,661 of them are indeed tappable, i.e., the view hierarchy attribute \texttt{clickable=True}, and 7,513 of them are not.

\subsubsection{How well can human users perceive the actual clickable state of an element as specified by developers or designers?}
To answer this question, we treat the \texttt{clickable} value of an element in the view hierarchy as the actual value and human labels as the predicted value for a precision and recall analysis. In this dataset of real mobile app screens, there were still many false signifiers for tappability potentially causing workers to misidentify tappable and not-tappable elements (see Table \ref{tab:accuracy_and_labels}). The workers labeled non-clickable elements as tappable 39\% of time. While the workers were significantly more precise in labeling clickable elements,  workers still marked clickable elements as not tappable 10\% of the time. The results were quite consistent across two rounds of data collection involving different workers and interface screens. These results further confirmed that tappability is an important usability issue worth investigation.

\subsection{Signifier Analysis}
To understand how users perceive tappability, we analyzed the potential signifiers affecting tappability in real mobile apps. These findings can help us understand human perception of tappability and help us build machine learning models to predict tappability. We investigated several visual and non-visual features based on previous understandings of common visual signifiers \cite{nngroup2015, affordancesAndDesign, materialDesign} and through exploration of the characteristics of the dataset. 

\begin{figure}
\centering
\includegraphics[width=\columnwidth]{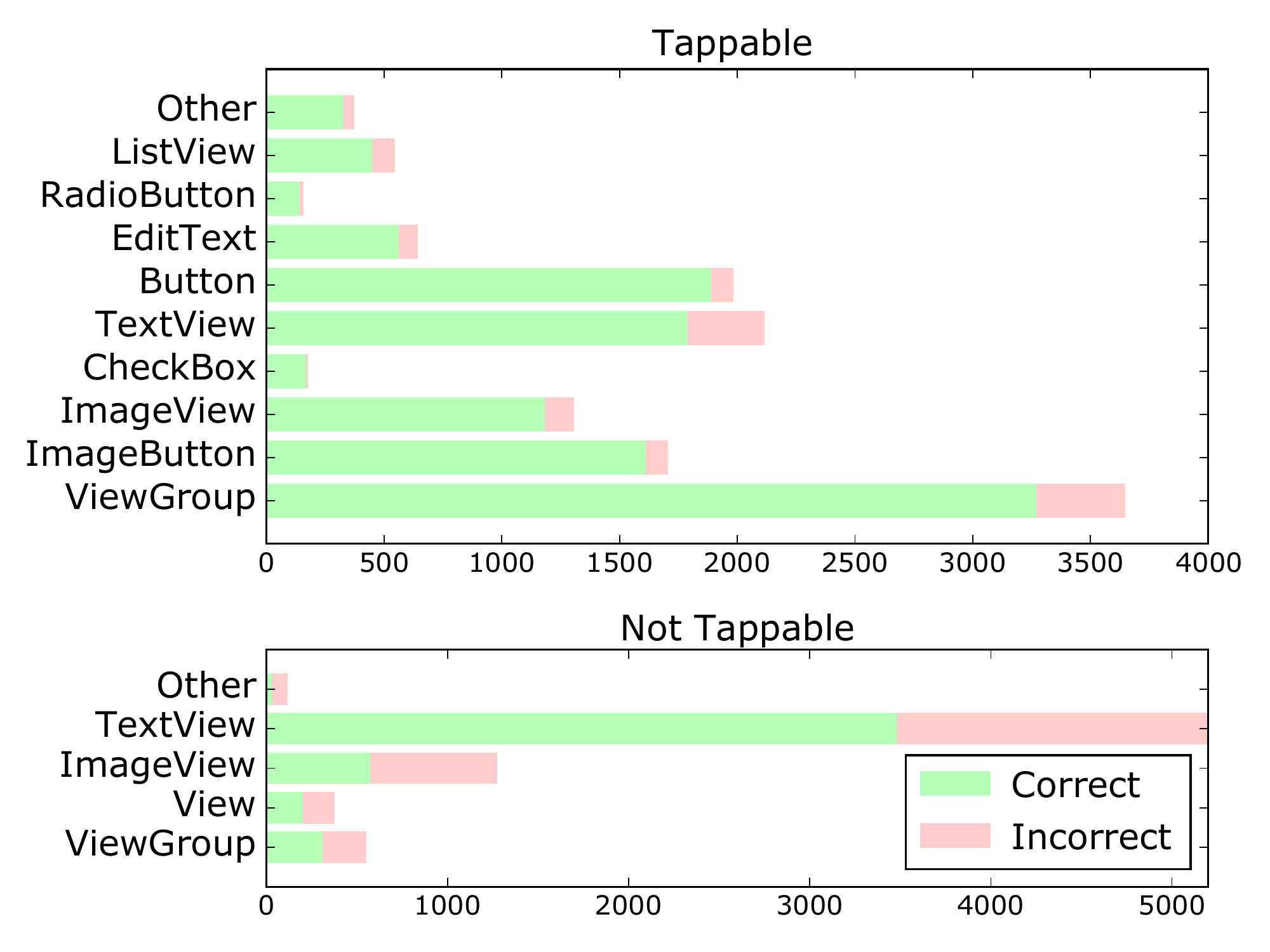}
    \caption{The number of tappable and not-tappable elements in several type categories with the bars colored by the relative amounts of correct and incorrect labels.}
    \Description{The graphic displays two horizontally arranged bar graphs where the bars correspond to different type categories of elements. The top bar graph is for tappable, and the bottom is for not tappable element types. The bars are colored according to the relative amounts of correct and incorrect labels within the type category. For tappable elements, ViewGroup and TextView types have higher proportions of incorrect labels (i.g., around 10-20\%). For not tappable elements, TextView has around 1/3 proportion of incorrect labels. ImageView, View, and ViewGroup have over 50\% incorrect labels.}
    \label{fig:accuracy_by_type}
    \vspace{-1em}
\end{figure}

\subsubsection{Element Type}
Several element types have conventions for visual appearance, thus users would consistently perceive them as tappable \cite{norman2013design} (e.g., buttons). We examined how accurately workers label each interface element type from a subset of Android class types in the Rico dataset \cite{deka2017rico}. Figure~\ref{fig:accuracy_by_type} shows the distribution of tappable and not-tappable elements by type labeled by human workers. Common tappable interface elements like Button and Checkbox did appear more frequently in the set of tappable elements. For each element type, we computed the accuracy by comparing the worker labels to the view hierarchy \texttt{clickable} values. For tappable elements, the workers achieved high accuracy for most types. For not-tappable elements, the two most common types, TextView and ImageView, had low accuracy percentages of only 67\% and 45\%, respectively. These interface types allow more flexibility in design than standard element types (e.g., RadioButton). Unconventional styles may make an element more prone to ambiguity in tappability. 

\begin{figure}
\centering
\includegraphics[width=\columnwidth]{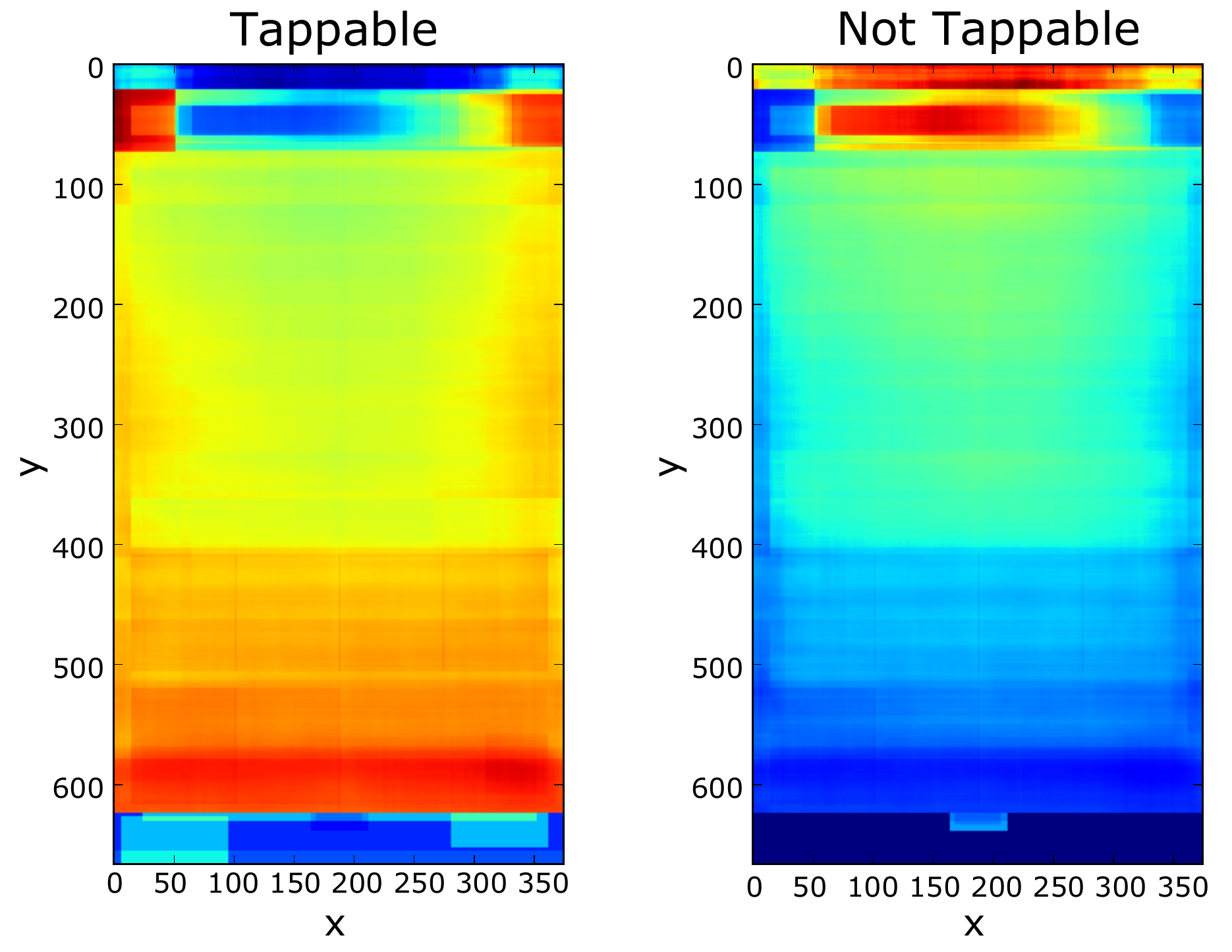}
    \caption{Heatmaps displaying the accuracy of tappable and not tappable elements by location where warmer colors represent areas of higher accuracy. Workers labeled not-tappable elements more accurately towards the upper center of the interface, and tappable elements towards the bottom center of the interface.}
    \Description{The image displays two rectangular heatmaps corresponding to the size of an app screen. The heatmap on the left is for tappable, and the right is for not tappable. The heatmaps display warmer colors (i.e., reds/yellows)  for high accuracy and colder colors (i.e., blues/greens) for low accuracy. The tappable heatmap shows high accuracy in the bottom area of the screen, and in the top right and left corner. The not tappable heatmap shows low accuracy in the bottom area of the screen and high accuracy in the top middle area of the screen and in the area corresponding to the header bar.}
    \label{fig:accuracy_by_location}
    \vspace{-1em}
\end{figure}

\subsubsection{Location}
We hypothesized that an element's location on the screen may have influenced the accuracy of workers in labeling its tappability. Figure \ref{fig:accuracy_by_location} displays a heatmap of the accuracy of the workers' labels by location. We created the heatmap by computing the accuracy per pixel, using the \texttt{clickable} attribute, across the 20,174 elements we collected using the bounding box of each element. Warm colors represent higher accuracy values. For tappable elements, workers were more accurate towards the bottom of the screen than the center top area. Placing a not-tappable element in these areas might confuse people. For tappable elements, there are two spots at the top region of high accuracy. We speculate that this is because these spots are where apps tend to place their Back and Forward buttons. For not-tappable elements, the workers were less accurate towards the screen bottom and highly accurate in the app header bar area with a corresponding area of low accuracy for tappable elements. This area is not tappable in many apps, so people may not realize any element placed there is tappable. 

\subsubsection{Size}
There was only a small difference in average size between labeled tappable and not-tappable elements. However, tappable elements labeled as not tappable were 1.9 times larger than tappable elements labeled as tappable indicating that elements with large sizes were more often seen as not tappable. Examining specific element types can reveal possible insights into why the workers may have labeled larger elements as not tappable. TextView elements tend to display labels but can also be tappable elements. From design recommendations, tappable elements should be labeled with short, actionable phrases \cite{tidwell2010designing}. The text labels of not-tappable TextView elements have an average and median size of 1.48 and 1.55 times larger respectively than those of tappable TextView elements. This gives us a hint that TextView elements may be following these recommendations. For ImageView elements, the average and median size for not-tappable elements were 2.39 and 3.58 times larger than for tappable elements. People may believe larger ImageView elements, typically displaying images, to be less likely tappable than smaller ImageView elements.


\begin{figure}
\centering
\includegraphics[width=\columnwidth]{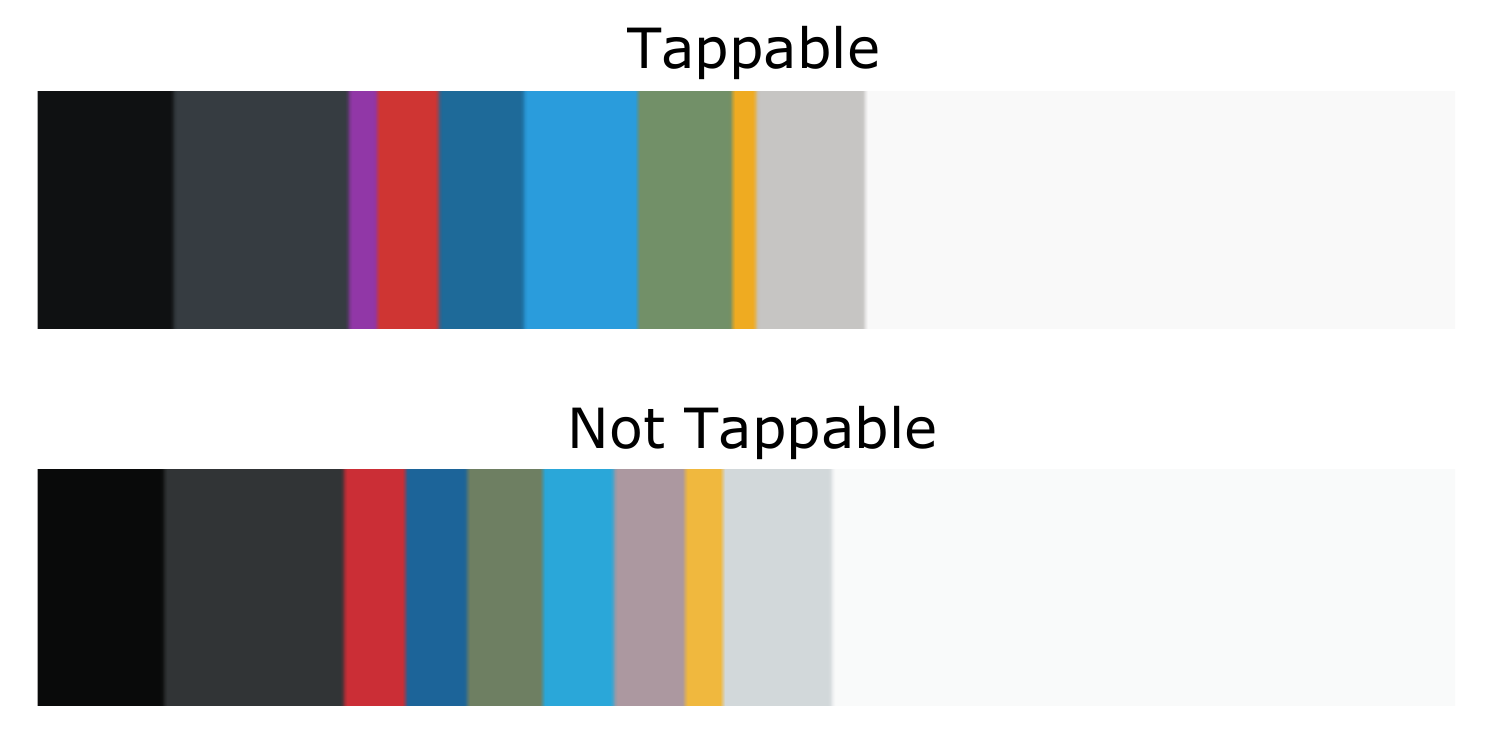}
    \caption{The aggregated RGB pixel colors of tappable and not-tappable elements clustered into the 10 most prominent colors using K-Means clustering.}
    \Description{The image shows two bars of a rectangular shape with sorted vertical bars within them of colors arranged from dark to light hues. The top of the image shows the tappable bar with sorted colors and the bottom shows the not tappable bar with sorted colors. The tappable bar has slightly wider bars for blue and bright colors and the not tappable bar has slightly wider bars for grey and light colors.}
    \label{fig:color_clustering}
    \vspace{-1em}
\end{figure}

\subsubsection{Color}
Based on design recommendations \cite{nngroup2015}, color can also be used to signify tappability. Figure ~\ref{fig:color_clustering} displays the top 10 dominant colors in each class of labeled tappable and not-tappable elements, which are computed using K-Means clustering. The dominant colors for each class do not necessarily denote the same set. The brighter colors such as blue and red have more presence, i.e., wider bars, in the pixel clusters for tappable elements than those for not-tappable ones. In contrast, not-tappable elements have more grey and white colors. We computed these clusters across the image pixels for 12 thousand tappable and 7 thousand not-tappable elements and scaled them by the proportion of elements in each set. These differences indicate that color is likely a useful distinguishing factor. 

\subsubsection{Words}

As not-tappable textual elements are often used to convey information, the number of words in these elements tend to be large. The mean number of words per element, based on the log-transformed word count in each element, was 1.84 times greater for not-tappable elements (Mean: 2.62, Median: 2) than tappable ones (Mean: 1.42, Median: 1). Additionally, the semantic content of an element's label may be a distinguishing factor based on design recommendations \cite{tidwell2010designing}.  We hypothesized that tappable elements would contain keywords indicating tappability, e.g., "Login". To test this,  we examined the top five keywords of tappable and not-tappable elements using TF-IDF analysis, with the set of words in all the tappable and not-tappable elements as two individual documents. The top 2 keywords extracted for tappable elements were "submit" and "close", which are common signifiers of actions. However, the remaining keywords for tappable elements, i.e., "brown", "grace" and "beauty", and the top five keywords for not-tappable elements, i.e., "wall", "accordance", "recently", "computer" and "trying", do not appear to be actionable signifiers. 
\section{Tappability Prediction Model}
Because it is expensive and time consuming to conduct user studies, it is desirable to develop automated techniques to examine the tappability of mobile interfaces. Although we can use the signifiers previously discussed as heuristics for this purpose, it would be difficult to manually combine them appropriately. It is also challenging to capture factors that are not obvious or hard to articulate. As such, we employed a deep learning approach to address the problem. Overall, our model is a feedforward neural network with a deep architecture (multiple hidden layers). It takes a concatenation of a range of features about the element and its screen and outputs a probability of how likely a human user would perceive an interface element as tappable.

\subsection{Feature Encoding}
Our model takes as input several features collected from the view hierarchy metadata and the screenshot pixel data of an interface. For each element under examination, our features include 1) semantics and functionality of the element, 2) the visual appearance of the element and the screen, and 3) the spatial context of the element on the screen.

\subsubsection{Semantic Features}
The length and the semantics of an element's text content are both potential tappability signifiers. For each element, we scan the text using OCR. To represent the semantics of the text, we use word embedding that is a standard way of mapping word tokens into a continuous dense vector that can be fed into a deep learning model. We encode each word token in an element as a 50-dimensional vector representation that is pre-learned from a Wikipedia corpus \cite{pennington2014glove}. When an element contains multiple words, we treat them as a bag of words and apply max pooling to their embedding vectors to acquire a single 50-dimensional vector as the semantic representation of the element. We also encode the number of word tokens each element contains as a scalar value normalized by an exponential function. 

\subsubsection{Type Features}
There are many standard element types that users have learned over time (e.g., buttons and checkboxes) \cite{norman2013design}. However, new element types are frequently introduced (e.g., floating action button). In our model, we include an element type feature as an indicator of the element's semantics. This feature allows the model to potentially account for these learned conventions as a users' background plays an important role in their decision. To encode the Type feature, we include a set of the 22 most common interface element types, e.g. TextView or Button. We represent the Type in the model as a 22-dimensional categorical feature, and collapse it into 6-dimensional embedding vector for training, which provides better performance over sparse input. Each type comes with a built-in or specified clickable attribute that is encoded as either 0 or 1. 

\begin{figure}
\centering
\includegraphics[width=0.9\columnwidth]{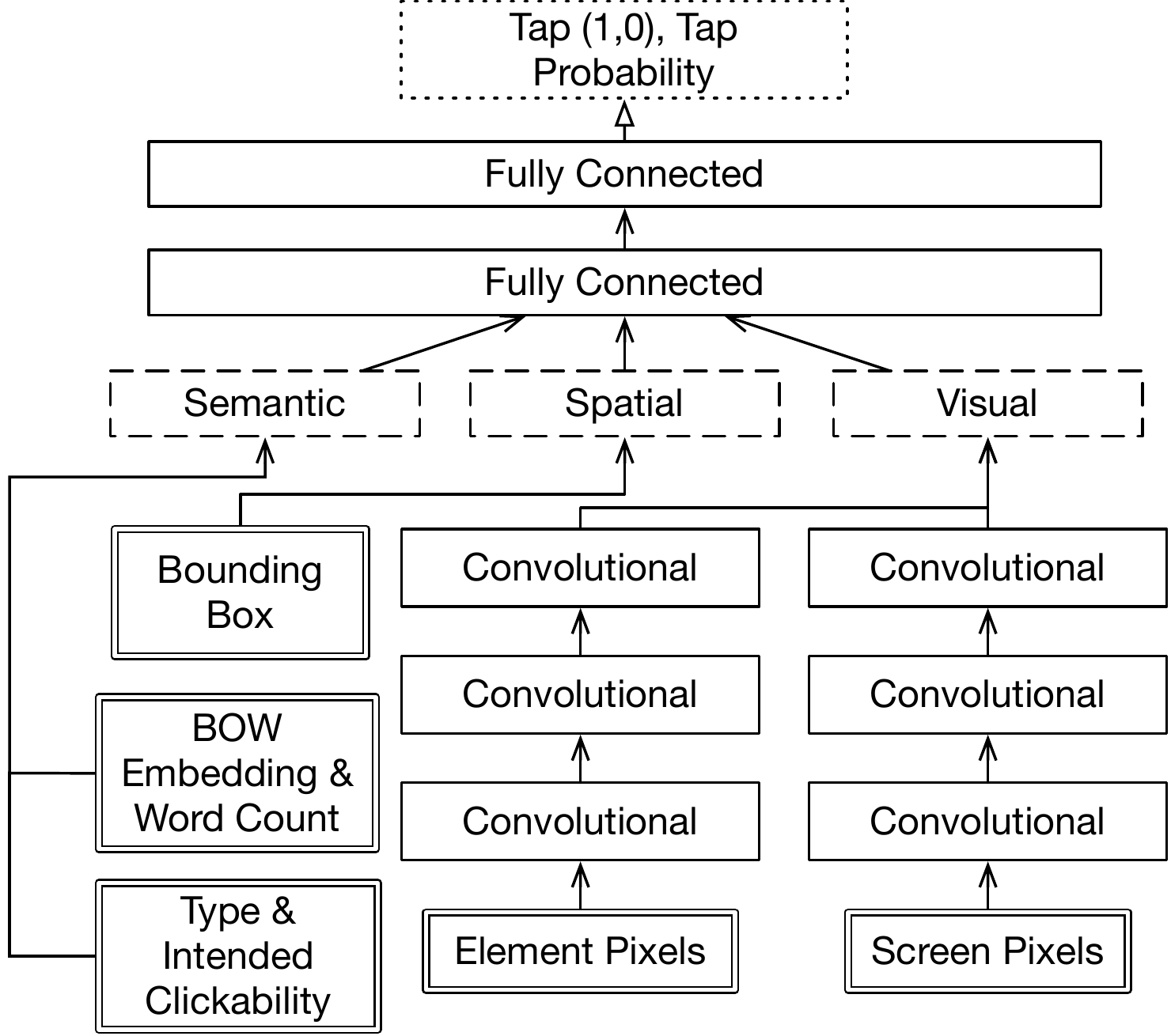}
    \caption{A deep model for tappability leverages semantic, spatial and visual features.}
    \Description{The figure displays the deep learning model architecture for the tappability prediction model in a reverse tree-like structure. The bottom of the figure has boxes corresponding to semantic, spatial, and visual features. The element pixel and screen pixel feature boxes have arrows pointing to two separate sets of 3 boxes with arrows between them corresponding to the convolutional layers. The type, and word features have arrows pointing upwards to a box labeled with "semantic". The bounding box has an arrow going upwards to a box labeled "spatial". The two sets of convolutional layers have arrows going upwards to a box labeled "visual". The three boxes "semantic", "spatial", and "visual" all three have arrows pointing to a "fully connected" box showing that the data from these features is being passed into a series of two fully connected layers, also represented by boxes. Finally, coming out of the top fully connected layer box, is an arrow pointing to the prediction box labeled "Tap Probability" where the model outputs a prediction of 1 or 0.}
    \label{fig:model_architecture}
    \vspace{-1em}
\end{figure}

\subsubsection{Visual Features}
As previously discussed, visual design signifiers such as color distribution can help distinguish an element's tappability. It is difficult to articulate the visual perception that might come into play and realize it as an executable rule. As a result, we feed an element's raw pixel values and the screen to which the element belongs to the network, through convolutional layers---a popular method for image processing. We resize the pixels of each element and format them as a 3D matrix in the shape of 32x32x3 where the height and width are 32, and 3 is the number of RGB channels. Contextual factors on the screen may affect the human's perception of tappability. To capture the context, we resize and format the entire screen as another visual feature. This manifests as a 3D matrix in the shape of 300x168x3 and preserves the original aspect ratio. As we will discuss later, a screen contains useful information for predicting an element's tappability even though such information is not easy to articulate.

\subsubsection{Spatial Features}
As location and size can be signifiers of tappability, we include them as features. We capture the element's bounding box as four scalar values: \textit{x}, \textit{y}, \textit{width}, and \textit{height}, and scale each value to the range of 0 and 1 by normalizing them using the screen width and height.

\subsection{Model Architecture \& Learning}
Figure~\ref{fig:model_architecture} illustrates our model architecture. To process the element and screenshot pixels, our network has three convolutional layers with ReLU \cite{nair2010rectified} activation. Each convolutional layer applies a series of 8 3x3 filters to the image to help the model progressively create a feature map. Each convolutional layer is followed by a 2x2 max pooling layer to reduce the dimensionality of the image data for processing. Finally, the output of the image layers is concatenated with the rest of the features into a series of two fully connected 100-dimensional dense layers using ReLU \cite{nair2010rectified} as the activation function. The output layer produces a binary classification of an element's tappability using a sigmoid activation function to transform the output into probabilities from zero to one. The probability indicates how likely the user would perceive the element as tappable. We trained the model by minimizing the sigmoid cross-entropy loss between the predicted values and the binary human labels on tappability of each element in the training data. For loss minimization, we used the Ada adaptive gradient descent optimizer with a learning rate of 0.01 and a batch size of 64. To avoid model overfitting, we applied a dropout ratio of 40\% to each fully connected layer to regularize the learning. We built our model using Tensorflow \cite{abadi2016tensorflow} in Python and trained it on a Tesla V100 GPU.

\subsection{Model Performance Results}
We evaluated our model using 10-fold cross validation with the crowdsourced dataset. In each fold, we used 90\% of the data for training and 10\% for validation, and trained our model for 100,000 iterations. Similar to an information retrieval task, we examine how well our model can correctly retrieve elements that users would perceive as tappable. We select an optimal threshold based on Precision-Recall AUC. Our model achieved a mean precision and recall, across the 10 folds of the experiment, of 90.2\% (SD: 0.3\%) and 87.0\% (SD: 1.6\%). To understand what these numbers imply, we analyzed how well the \texttt{clickable} attribute in the view hierarchy predicts user tappability perception: precision 89.9\% (SD: 0.6\%) and recall 79.6\% (SD: 0.8\%). While our model has a minor improvement on precision, it outperforms the \texttt{clickable} attribute on recall considerably by over 7\%. 

Although identifying not-tappable elements is less important in real scenarios, to better understand the model, we report the performance concerning not-tappable elements as the target class. Our model achieved a mean precision 70\% (SD: 2\%) and recall 78\% (SD: 3\%), which improves precision by 9\%, with a similar recall, over the \texttt{clickable} attribute (precision 61\%, SD: 1\% and recall 78\%, SD: 2\%). One potential reason that not-tappable elements have a relatively low accuracy is that they tend to be more diverse, leading to more variance in the data. 

In addition, our original dataset had an uneven number of tappable and not-tappable elements (14,301 versus 5,871), likely causing our model to achieve higher precision and recall for tappable elements than not-tappable ones. Therefore we created a balanced dataset by upsampling the minority class (i.e., not-tappable). On the balanced dataset, our model achieved a mean precision and recall of 82\% and 84\% for identifying tappable elements, and a mean precision and recall of 81\% and 86\% for not-tappable elements. Table~\ref{tab:confusion_matrix} shows the confusion matrix for the balanced dataset. Compared to using view hierarchy \texttt{clickable} attribute alone, which achieved mean precision 79\% and recall 80\% for predicting tappable elements, and 79\% and 78\% for not-tappable ones, our model is consistently more accurate across all the metrics. These performance improvements show that our model can effectively help developers or designers identify tappability misperceptions in their mobile interfaces.

\begin{table}
\centering
\begin{tabular}{c | c | c|}
\cline{2-3}
 & \vtop{\hbox{\strut \textbf{Predicted}} \hbox{\strut \textbf{Tappable}}} & \vtop{\hbox{\strut \textbf{Predicted}} \hbox{\strut \textbf{Not Tappable}}} \\
\hline
\multicolumn{1}{|c|}{\textbf{Actual Tappable}} & 1195 & 260  \\
\hline
\multicolumn{1}{|c|}{\textbf{Actual Not Tappable}} & 235 & 1170 \\
\hline
\end{tabular}
\vspace{1em}
\caption{Confusion matrix for the balanced dataset, averaged across the 10 cross-validation experiments.}~\label{tab:confusion_matrix}
\vspace{-2em}
\end{table}

\subsection{Human Consistency \& Model Behaviors}
We speculate that our model did not achieve even higher accuracy because human perception of tappability can be inherently inconsistent as people have their own experience in using and learning different sets of mobile apps. This can make it challenging for the model to achieve perfect accuracy. To examine our hypothesis, we collected another dataset via crowdsourcing using the same interface as shown in Figure~\ref{fig:tappability_task}. We selected 334 screens from the Rico dataset, which were not used in our previous rounds of data collection. We recruited 290 workers to perform the same task of marking each selected element as either tappable or not tappable. However, each element was labeled by 5 different workers to enable us to see how much these workers agree on the tappability of an element. In total, there were 2,000 unique interface elements and each was labeled 5 times. In total, 1,163 elements (58\%) were entirely consistent among all 5 workers which include both tappable and not-tappable elements. We report two metrics to analyze the consistency of the data statistically.  The first is in terms of an agreement score \cite{wobbrock2005maximizing} that is computed using the following formula: 

\begin{equation}
\label{eq:guess}
A = \frac{\sum\limits_{e\in E}\sum\limits_{r\in R} \left (  \frac{\left | R_{i} \right | }{\left | R_{e} \right |} \right )^{2}}{\left | E \right |} \times 100\%
\end{equation}

Here, $e$ is an element in the set of all interface elements $E$ that were rated by the workers, $R_{e}$ is the set of ratings for an interface element $e$, and $R_{i}$ is the set of ratings in a single category (0: not tappable, 1: tappable). We also report the consistency of the data using Fleiss' Kappa \cite{fleiss1971measuring}, a standard inter-rater reliability measure for the agreement between a fixed number of raters assigning categorical ratings to items. This measure is useful because it computes the degree of agreement over what would be expected by chance. As there are only two categories, the agreement by chance is high. The overall agreement score across all the elements using Equation \ref{eq:guess} is \textit{0.8343}. The number of raters is 5 for each element on a screen, and across 334 screens, resulting in an overall Fleiss' Kappa value of 0.520 \textit{(SD=0.597, 95\% CI [0.575,0.618], P=0)}. This corresponds to a "Moderate" level agreement according to \cite{landis1977application}. What these results demonstrate is that while there is a significant amount of consistency in the data, there still exists a certain level of disagreement on what elements are tappable versus not tappable. Particularly, consistency varies across element \textit{Type} categories. For example, \textit{View} and \textit{ImageView} elements were labeled far less consistently (\textit{0.52, 0.63}) than commonplace tappable element types such as \textit{Button} (94\%), \textit{Toolbar} (100\%), and \textit{CheckBox} (95\%). View and ImageView elements have more flexibility in design, which may lead to more disagreement.

\begin{figure}
\centering
\includegraphics[width=\columnwidth]{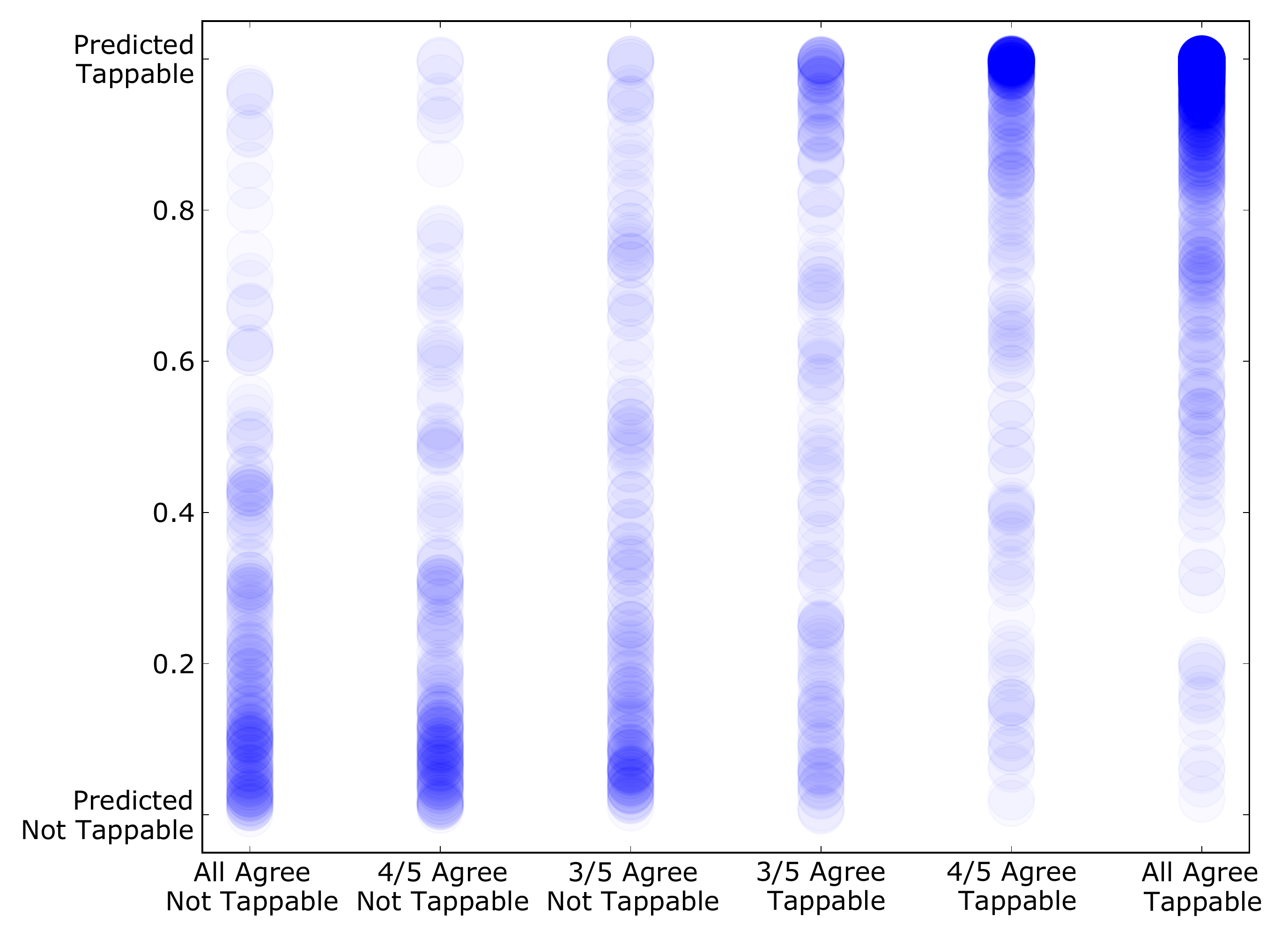}
    \caption{The scatterplot of the tappability probability output by the model (the Y axis) versus the consistency in the human worker labels (the X axis) for each element in the consistency dataset.}
    \Description{The figure displays a scatterplot with the vertical axis having the labels of 0.0 - Predicted Not Tappable, 0.2, 0.4, 0.8, and Predicted Tappable. The horizontal axis has the labels from left to right of "All Agree Not Tappable", "4/5 Agree Not Tappable", "3/5 Agree Not Tappable", "3/5 Agree Tappable", "4/5 Agree Tappable", and "All Agree Tappable". There is vertical line of blue dots above each horizontal marker. The three lines representing majority not tappable labels  by workers have darker areas towards the bottom meaning there are more prediction dots appearing at the bottom of the vertical axis corresponding to a not tappable prediction. The "3/5 Agree Tappable" line has evenly distributed dots across the vertical axis. The "4/5 Agree Tappable" and "All Agree Tappable" have much darker areas towards the top of the vertical line extending to the vertical marker of "Predicted Tappable" meaning that there are many dots overlapping towards the top of the vertical axis.}
    \label{fig:consistency_scatterplot}
    \vspace{-1em}
\end{figure}

To understand how our model predicts elements with ambiguous tappability, we test our previously trained model on this new dataset. Our model matches the uncertainty in human perception of tappability surprisingly well (see Figure~\ref{fig:consistency_scatterplot}). When workers are consistent on an element's tappability (two ends on the X axis), our model tends to give a more definite answer---a probability close to 1 for tappable and close to 0 for not tappable. When workers are less consistent on an element (towards the middle of the X axis), our model predicts a probability closer to 0.5.

\subsection{Usefulness of Individual Features}
One motivation to use deep learning is to alleviate the need for extensive feature engineering. Recall that we feed the entire screenshot of an interface to the model to capture contextual factors affecting the user's decision that can not be easily articulated. Without the screenshot pixels as input, there is a noticeable drop in precision and recall for tappable of 3\% and 1\%, and for not-tappable, an 8\% drop in precision but no change in recall. This indicates that there is useful contextual information in the screenshot affecting the users' decisions on tappability. We also examined removing the \textit{Type} feature from the model, and found a slight drop in precision about 1\% but no change in recall for identifying tappable elements. The performance change is similar for the not-tappable case with 1.8\% drop in precision and no drop in recall. We speculate that removing the Type feature only caused a minor impact likely because our model has captured some of element type information through its pixels.

\begin{figure*}[t]
\centering
\includegraphics[width=\textwidth]{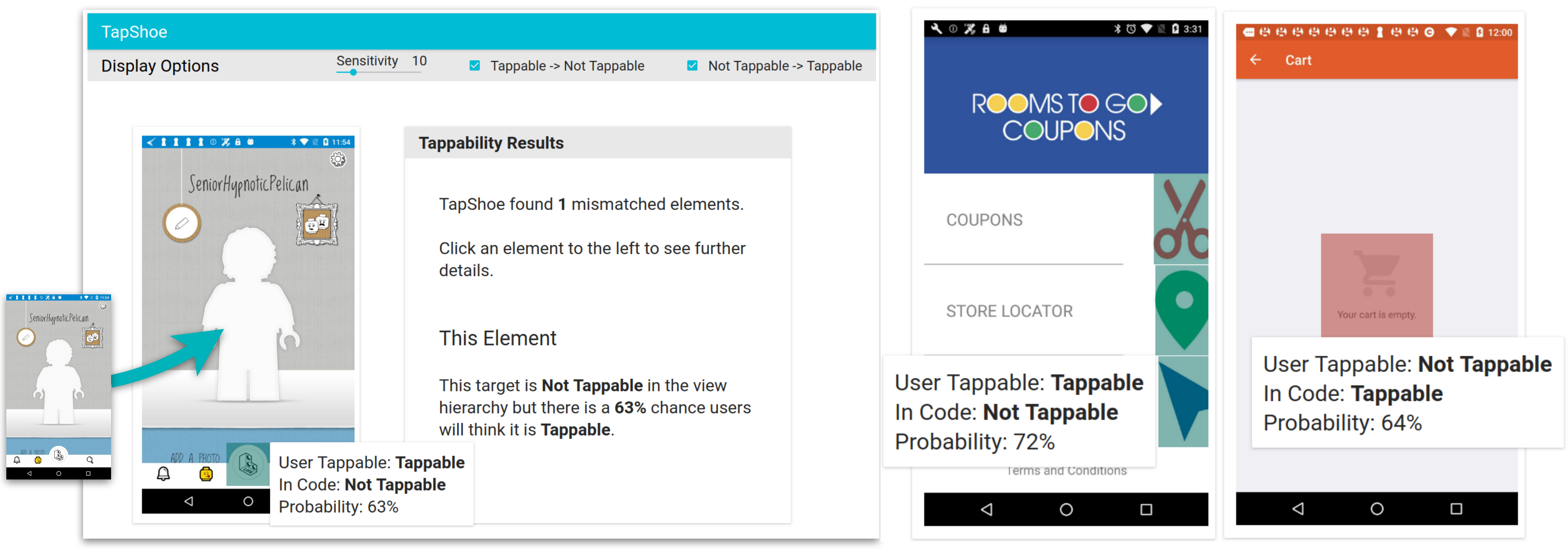}
    \caption{The TapShoe interface. An app designer drag and drops a UI screen on the left. TapShoe highlights interface elements whose predicted tappability is different from its actual tappable state as specified in its view hierarchy.}
    \Description{The left hand side of the image shows a screenshot being dragged into a hotspot area on the TapShoe interface. The interface has the screenshot on the left, and the tappability analysis results on the right. The image displays a highlighted element on the screenshot with results active showing the tappability prediction probability for an element and its tappable label. The element is a circular element at the bottom of the screen and has been predicted as tappable (63\%), yet in code is not tappable. 
    
    The tappability results area contains a more detailed description of the results for the selected element. On the far right side of the figure, there are two screenshots with two different elements highlighted, and tooltips active showing their predictions. The left screenshot has a colorful, large arrow icon highlighted with a tappable prediction (72\%), yet in code it is not tappable. The right screenshot has a large shopping cart icon in the center of the screen with a not tappable prediction (64\%), yet in code the element is tappable.}
    \label{fig:tap_shoe_interface}
    \vspace{-1em}
\end{figure*}

\section{TapShoe Interface}
We created a web interface for our tappability model called TapShoe (see Figure~\ref{fig:tap_shoe_interface}). The interface is a proof-of-concept tool to help app designers and developers examine their UI's tappability. We describe the TapShoe interface from the perspective of an app designer, Zoey, who is designing an app for deal shopping, shown in the right hand side of Figure~\ref{fig:tap_shoe_interface}. Zoey has redesigned some icons to be more colorful on the home page links for "Coupons", "Store Locator", and "Shopping". Zoey wants to understand how the changes she has made would affect the users' perception of which elements in her app are tappable. First, Zoey uploads a screenshot image along its view hierarchy for her app by dragging and dropping them into the left hand side of the TapShoe interface. Once Zoey drops her screenshot and view hierarchy, TapShoe analyzes her interface elements, and returns a tappable or not-tappable prediction for each element. The TapShoe interface highlights the interface elements with a tappable state, as specified by Zoey in the view hierarchy, that does not match up with user perception as predicted by the model. 

Zoey sees that the TapShoe interface highlighted the three colorful icons she redesigned. These icons were not tappable in her app but TapShoe predicted that the users would perceive them as tappable. She examines the probability scores for each element by clicking on the green hotspots on the screenshot to see informational tooltips. She adjusts the sensitivity slider to change the threshold for the model's prediction. Now, she sees that the "Coupons" and "Store Locator" icon are not highlighted and that the arrow icon has the highest probability of being perceived as tappable. She decides to make all three colorful icon elements interactive and extend the tappable area next to "Coupons", "Store Locator", and "Website". These fixes prevent her users from the frustration of tapping on these elements with no response. 

We implemented the TapShoe interface as a web application (JavaScript) with a Python web server. The web client accepts an image and a JSON view hierarchy to locate interface elements. The web server queries a trained model, hosted via a Docker container with the Tensorflow model serving API, to retrieve the predictions for each element. 
\section{Informal Feedback from Designers}
To understand how the TapShoe interface and tappability model would be useful in a real design context, we conducted informal design walkthroughs with 7 professional interface designers at a large technology company. The designers worked on design teams for three different products. We demonstrated TapShoe to them and collected informal feedback on the idea of getting predictions from the tappability model, and on the TapShoe interface for helping app designers identify tappability mismatches. We also asked them to envision new ways they could use the tappability prediction model beyond the functionality of the TapShoe interface. The designers responded positively to the use of the tappability model and TapShoe interface, and gave several directions to improve the tool. Particularly, the following themes have emerged.

\subsection{Visualizing Probabilities}
The designers saw high potential in being able to get a tappability probability score for their interface elements. Currently, the TapShoe interface displays only probabilities for elements with a mismatch based on the threshold set by the sensitivity slider. However, several of the designers mentioned that they would want to see the scores for all the elements. This could give them a quick glance at the tappability of their designs as a whole. Presenting this information in a heatmap that adjusts the colors based on the tappability scores could help them compare the relative level of tappability of each element. This would allow them to deeply examine and compare interface elements for which tappability signifiers are having an impact. The designers also mentioned that sometimes, they do not necessarily aim for tappability to be completely binary. Tappability could be aimed to be higher or lower along a continuous scale depending on an element's importance. In an interface with a primary action and a secondary action, they would be more concerned that people perceive the primary action as tappable than the secondary action.

\subsection{Exploring Variations}
The designers also pointed out the potential of the tappability model for helping them systematically explore variations. TapShoe's interface only allows a designer to upload a single screen. However, the designers envisioned an interface to allow them to upload and compare multiple versions of their designs to systematically change signifiers and observe how they impact the model's prediction. This could help them discover new design principles to make interface elements look more or less tappable. It could also help them compare more granular changes at an element level, such as different versions of a button design. As context within a design can also affect an element's tappability, they would want to move elements around and change contextual design attributes to have a more thorough understanding of how context affects tappability. Currently, the only way for them to have this information is to conduct a large tappability study, which limits them to trying out only a few design changes at a time. Having the tappability model output could greatly expand their current capabilities for exploring design changes that may affect tappability. 

\subsection{Model Extension and Accuracy}
Several designers wondered whether the model could extend to other platforms. For example, their design for desktop or web interfaces could benefit from this type of model. Additionally, they have collected data that our model could already use for training. We believe our model could help them in this case as it would be simple to extend to other platforms or to use existing tappability data for training.

We also asked the designers about how they feel about the accuracy of our model. The designers already thought that the model could be useful in its current state even for helping them understand the relative tappability of different elements. Providing a confidence interval for the prediction could aid in giving them more trust in the prediction.
\section{Discussion}
Our model achieves good accuracy at predicting tappable and not-tappable interface elements and the TapShoe tool and model are well-received by designers. Here we discuss the limitations and directions for future work. 

One limitation is that our TapShoe interface, as a proof-of-concept, demonstrates one of many potential uses for the tappability model. We intend to build a more complete design analytics tool based on designers' suggestions, and conduct further studies of the tool by following its use in a real design project. Particularly, we will update the TapShoe interface to take early stage mockups other than UI screens that are equipped with a view hierarchy. This is possible because designer can mark up elements to be examined in a mockup without having to implement it. 

Our tappability model is only trained on Android interfaces and therefore the results may not generalize well to other platforms. However, our model relies on general features available in many UI platforms (e.g., element bounding boxes and types). It would be entirely feasible to collect a similar dataset for different platforms to train our model and the cost for crowdsourcing labeling is relatively small. In fact, we can apply a similar approach to new UI styles that involve drastically different design concepts, e.g., emerging UI styles in AR/VR. 

From our consistency evaluation, we learned that people's perception of tappability is not always consistent. In the future, we plan to explore ways to improve the model's performance with inconsistent data. These methods could extend our tappability annotation task beyond a simple binary rating of tappable versus not-tappable to a rating that incorporates uncertainty, e.g., adding a "Not sure" option or a scale of confidence in labels.

The tappability model that we developed is a first step towards modeling tappability. There may potentially be other features that could add predictive power to the model. As we begin to understand more of the features that people use to determine which elements are tappable and not tappable, we can incorporate these new features into a deep learning model as long as they are manifested in the data. For example, we used the Type feature as a way to account for learned conventions, i.e., the behavior that users have learned over time. As users are not making a tappable decision solely based on the visual properties of the current screen, we intend to explore more features that can capture user background.

Lastly, identifying the reasons behind tappable or not-tappable perception could potentially enable us to offer recommendations for a fix. This also requires us to communicate these reasons with the designer in a human-understandable fashion. There are two approaches to pursue this. One is to analyze how the model relies on each feature, although understanding the behavior of a deep learning model is challenging and it is an active area in the deep learning field. The other approach is to train the model to recognize the human reasoning behind their selection. Progress in this direction will allow a tool to provide a more complete and useful output to the designers.

\section{Conclusions}
We present an approach to model interface tappability at scale. We collected a large dataset of tappability examples via crowdsourcing and analyzed a variety of tappability signifiers based on the dataset. We then designed and trained a deep model that achieved reasonable accuracy in predicting human perception on tappability. We analyzed the model behavior in relation to uncertainty in human tappability perception. Finally, we buit TapShoe, a tool that uses the deep model to examine interface tappability, which received positive feedback from 7 professional interaction designers who saw its potential as a useful tool for their real design projects.

\bibliographystyle{ACM-Reference-Format}
\balance
\bibliography{references}
\end{document}